\documentclass[a4paper,onecolumn]{article}
\usepackage{amsfonts}
\usepackage{graphicx}
\usepackage{amsmath}

\input{tcilatex}
\begin{document}

\title{From periodic sampling to irregular sampling through PNS (Periodic
Nonuniform Sampling)}
\author{Bernard Lacaze \and T\'{e}SA, 7 bd de la Gare, 31500 Toulouse \and %
bernard.lacaze@tesa.prd.fr}
\maketitle

\begin{abstract}
Resampling is an operation costly in calculation time and accuracy. It
regularizes irregular sampling, replacing $N$ data by $N$ periodic
estimations. This stage can be suppressed, using formulas built with
incoming data and completed by sequences of elements which influence
decreases when the number of data increases. Obviously, some spectral
properties (for processes) and some asymptotic properties (for the sampling
sequence) have to be fulfilled. In this paper, we explain that it is
possible to elaborate a logical theory, starting from the ordinary periodic
sampling, and generalized by the PNS (Periodic Nonuniform Sampling), \ to
treat more general irregular samplings. The "baseband spectrum" hypothesis
linked to the "Nyquist bound" (or Shannon bound) is generalized to spectra
in a finite number of intervals, suited to the "Landau condition".

\textit{keywords : }periodic sampling, periodic nonuniform sampling,
irregular sampling, Vostok ice core.
\end{abstract}

\section{Introduction}

In signal theory, sampling addresses a time sequence \textbf{t}=$\left\{
t_{n},n\in \mathcal{A}\right\} $ and values $g\left( t_{n}\right) $ of some
function $g\left( t\right) ,$ belonging to a given class $\mathcal{B}$. We
look for approximations $\widetilde{g}\left( t\right) $ close to $g\left(
t\right) $ from the set of $g\left( t_{n}\right) ,n\in \mathcal{A}$. Among
possibilities, we have linear combinations such as%
\begin{equation}
\widetilde{g}\left( t\right) =\sum_{k\in \mathbb{Z}}a_{k}\left( t\right)
g\left( t_{k}\right) .
\end{equation}%
Coefficients $a_{k}\left( t\right) $ characterize the class $\mathcal{B}$. $%
\widetilde{g}\left( t\right) $ may be equal to $g\left( t\right) $ or not,
and may minimize some given distance. Practically, formula $\left( 1\right) $
will be truncated (apodized) when $\mathcal{A}$ is not finite, to provide
usable results.

The sampling sequence \textbf{t=} $\left\{ nT,n\in \mathbb{Z}\right\} $
defines the simplest "periodic sampling". Bibliography on this basic
situation goes back to A. L. Cauchy and E. Borel \cite{Bore}, \cite{Laca10}.
The "sampling formula" of Shannon or Nyquist (or others) refers to functions
with Fourier transform cancelling outside the interval $\left(
-1/2T,1/2T\right) $. As explained below, the problem is closely linked to
usual FSD (Fourier Series Developments). PNS$N$ (Periodic Nonuniform
Sampling of order $N$) are concatenations of $N$ periodic samplings of same
period. They were introduced in the 1950s, firstly as variations on the
purely periodic case \cite{Yen}, \cite{Kohl}. They allow to bypass aliasing
problems, to simplify and to produce cheaper electronical devices. The
particular case of Coset (or Multi-Coset) assumes the existence of a primary
clock. Cosets are produced from subsequences of same period \cite{Yen}, \cite%
{Yuan}, \cite{Venk}. They used in the domain of communications, but
algorithms can be disputed.

Given some function $g\left( t\right) ,$ we define the set $g_{n}\left(
t\right) ,n=1,2,..,N,$ as outputs of bandpass filters on disjoined sets $%
\Delta _{n},$ where $\Delta =\cup _{n=1}^{N}\Delta _{n}$ is the "spectral
support" of $g\left( t\right) .$ For a large collection of $\Delta _{n},$ we
explain that a PNS$N$ is able to recover the $g_{n}\left( t\right) .$
Extending sampling properties from ordinary functions to random processes is
very easy. In both cases, we have to develop the same exponential functions,
but with convergence criteria which can be different. Therefore, we prove
that PNS$N,$ when $N$ is large enough, is able to determine bandpass
filterings and power spectra of stationary processes. Particularly, ice
cores provide a wide scope of data at irregular times and allow to verify
improvements thanks to methods displayed in this paper.

This paper is organized as follows. Section 2 explains the periodic sampling
as dependent of spectral supports. Section 3 develops the PNS$N$ for
particular partitions of spectra. Subsequent sections extend these results
for more general samplings, and section 8 specifically addresses data of the
Vostok ice score which is a basis for the study of the last million year
climate.

\section{Periodic sampling}

\subsection{Shannon formula}

1) We consider real or complex functions $g\left( t\right) $ such that

\begin{equation}
g\left( t\right) =\int_{-1/2T}^{1/2T}G\left( f\right) e^{2i\pi ft}df
\end{equation}%
with $G\left( f\right) =0,f\notin \left( -1/2T,1/2T\right) .$ $g\left(
t\right) $ is the Fourier transform (FT) of $G\left( f\right) ,$ and $\left(
-1/2T,1/2T\right) $ is a support of $G\left( f\right) .$ We assume that the
bounds of used sets (open, closed..) have no incidence on computations. The
historical "sampling formula" addresses the sampling sequence \textbf{t}=$%
\left\{ nT,n\in \mathbb{Z}\right\} .$

From $\left( 2\right) $ we have

\begin{equation}
g\left( nT\right) =\int_{-1/2T}^{1/2T}G\left( f\right) e^{2i\pi fnT}df.
\end{equation}%
To "develop" $g\left( t\right) $ as a function of the $g\left( nT\right) $
is equivalent to develop $e^{2i\pi ft}$ as a function of the $e^{2i\pi fnT}$
when $f\in \left( -1/2T,1/2T\right) .$ We know that, whatever the real $t$

\begin{equation}
e^{2i\pi ft}=\sum_{n=-\infty }^{\infty }\text{sinc}\left[ \pi \left( \frac{t%
}{T}-n\right) \right] e^{2i\pi fnT},-1/2T<f<1/2T
\end{equation}%
where sinc$x=\left( \sin x\right) /x$ is the "sinc function" (see appendix 1
about the Fourier series). In $\left( 2\right) ,$ we replace $e^{2i\pi ft}$
by the rhs of $\left( 4\right) ,$ we change the order of summations
(assuming that this operation is allowed), to obtain

\begin{equation}
g\left( t\right) =\sum_{n=-\infty }^{\infty }\text{sinc}\left[ \pi \left( 
\frac{t}{T}-n\right) \right] g\left( nT\right)
\end{equation}%
It is the "sampling formula" of Shannon, Nyquist and others. Conversely, $%
\left( 5\right) $ does not work if the "support "of $G\left( f\right) $ is
not included in $\left( -1/2T,1/2T\right) .$ $1/T,$ the length of this
interval, is the maximum which is compatible with the "sampling period" $T.$
It is the "Nyquist bound".

2) From $\left( 3\right) $ $g\left( nT\right) $ is the Fourier series
coefficient of the periodic function equal to $G\left( f\right) $ on the
period $\left( -1/2T,1/2T\right) .$ The Dirichlet theorem states that the
set of $g\left( nT\right) $ determines $G\left( f\right) $ and then $g\left(
t\right) $ (see appendix). We find in Emile Borel (\cite{Bore}, 1897) the
following sentences:

"Let's put

\begin{equation*}
g\left( z\right) =\int_{-\pi }^{\pi }\phi \left( x\right) e^{izx}dx,
\end{equation*}%
and let's assume that $\phi \left( x\right) $ fulfills the Dirichlet's
conditions. Therefore, if we know the values of $g\left( z\right) $ for $%
z=0,\pm 1,\pm 2,...$ the function $\phi \left( x\right) $ is determined and
consequently the entire function $g\left( z\right) $ is \textit{known
without ambiguousness"} (\textit{italics} come from Borel).

3) Formula $\left( 5\right) $ is verified in the same time as $\left(
2\right) ,$ assuming that a reversal of summations is possible. Using $%
\left( 4\right) ,$ we see that $\left( 5\right) $ is true for the larger set
of functions as

\begin{equation}
g\left( t\right) =\int_{-1/2T}^{1/2T}G\left( f\right) e^{2i\pi
ft}df+\sum_{k}a_{k}e^{2i\pi f_{k}t}
\end{equation}%
where $f_{k}\in \left( -1/2T,1/2T\right) $ $\ ($at least when the number of $%
f_{k}$ is finite). From a "mechanical" point of view, $\left( 6\right) $
isolates "masses" distributed with a density $G\left( f\right) $ among
"masses" located at points $f_{k}.$ For acoustic specialists and others, we
have a mix of a "noise" (with spectral density $G\left( f\right) )$ and of a
"signal" made of pure frequencies $f_{k}.$

Extracting pure frequencies (or concentrations of power) is a problem of
"signal theory". Therefore, decomposition $\left( 6\right) $ is natural,
because it separates things of different nature. Curiously, the community
prefers both components ("continuous" and "discrete") under the shape $%
\left( 2\right) .$

In this form, the rhs $\left( 2\right) $ is no longer an ordinary integral
(for instance a Riemann integral). We consider a new set of mathematical
items, the "Dirac functions" $\delta \left( f\right) $ which verify for $%
f_{k}\in \left( -1/2T,1/2T\right) ,$ and by convention 
\begin{equation}
e^{2i\pi f_{k}t}=\int_{-1/2T}^{1/2T}\delta \left( f-f_{k}\right) e^{2i\pi
ft}df.
\end{equation}%
In this way, the last term in $\left( 6\right) $ enters the integral (it is
a shoehorn procedure). $G\left( f\right) $ is replaced by

\begin{equation}
G\left( f\right) +\sum_{k}a_{k}\delta \left( f-f_{k}\right) .
\end{equation}%
In this form, $G\left( f\right) $ is the continuous part of the FT (it is a
density) and the residual is the discrete part ("the spectral lines").

\subsection{The general case}

1) Instead of $\left( 2\right) ,$ we consider the more general shape

\begin{equation}
g\left( t\right) =\int_{\Delta }G\left( f\right) e^{2i\pi ft}df
\end{equation}%
where $\Delta $ is for instance a union of intervals (it is a $G\left(
f\right) $ support). Using the partition $\mathbb{R}$=$\cup _{k\in \mathbb{Z}%
}\left( \frac{k}{T}-\frac{1}{2T},\frac{k}{T}+\frac{1}{2T}\right) ,$ $\left(
9\right) $ becomes 
\begin{equation*}
g\left( t\right) =\int_{-1/2T}^{1/2T}e^{2i\pi ft}H_{t}\left( f\right) df
\end{equation*}%
\begin{equation}
H_{t}\left( f\right) =H_{t+T}\left( f\right) =\sum_{k\in \mathbb{Z}}G\left(
f+\frac{k}{T}\right) e^{2i\pi kt/T}.
\end{equation}%
Particularly%
\begin{equation*}
g\left( nT\right) =\int_{-1/2T}^{1/2T}e^{2i\pi fnT}H_{0}\left( f\right) df.
\end{equation*}%
$Tg\left( nT\right) $ \ is the Fourier coefficient of $H_{0}\left( f\right)
. $ Conversely%
\begin{equation*}
H_{0}\left( f\right) =T\sum_{k\in \mathbb{Z}}g\left( nT\right) e^{-2i\pi
fnT}.
\end{equation*}%
Knowing $g\left( nT\right) ,n\in \mathbb{Z}$, is equivalent to knowing 
\begin{equation}
H_{0}\left( f\right) =\sum_{k\in \mathbb{Z}}G\left( f+\frac{k}{T}\right)
\end{equation}%
and knowing of $G\left( f\right) $ implies $g\left( t\right) .$ To obtain $%
g\left( t\right) $ from $H_{0}\left( f\right) $ it is necessary that the rhs
of $\left( 11\right) $ contains at most one non-zero term. It is the
"alias-free" condition. Consequently, the length of $\Delta $ (the $G\left(
f\right) $ support) cannot be larger than $1/T,$ but can be arbitrary close
of $1/T,$ whatever the bounds of any interval which contain $\Delta .$
Knowing $\Delta $ allows to retrieve $g\left( t\right) $ if, $T$ being
given, the alias-free condition is fulfilled.

2) For instance, following formulas are fitted to simple $\Delta $ \ \ \ 
\begin{equation*}
\text{a) \ \ \ \ \ \ \ \ \ \ \ \ \ }\ \Delta =\left( a,a+\frac{1}{T}\right)
\Longleftrightarrow
\end{equation*}%
\begin{equation*}
g\left( t\right) =e^{2i\pi t\left( a+1/2T\right) }\sum_{k\in \mathbb{Z}%
}\left( -1\right) ^{k}e^{-2i\pi kT}\text{sinc}\pi \left( \frac{t}{T}%
-k\right) g\left( kT\right) .
\end{equation*}%
This formula comes from the Fourier development of $e^{2i\pi ft}$ on $\Delta 
$ (see appendix). 
\begin{equation*}
\text{b) \ \ \ \ \ \ \ \ \ }\Delta =\left( \frac{-2k-1}{2T},\frac{-k}{T}%
\right) \cup \left( \frac{k}{T},\frac{2k+1}{2T}\right) ,k\in \mathbb{N}%
\Longleftrightarrow
\end{equation*}%
\begin{equation*}
g\left( t\right) =\frac{1}{\sin \pi t/T}\left[ -B\left( t\right) \sin 2\pi
kt/T+A\left( t\right) \sin \pi \left( 2k+1\right) t/T\right]
\end{equation*}%
\begin{equation*}
A\left( t\right) =\sum_{n\in \mathbb{Z}}\text{sinc}\pi \left( \frac{t}{T}%
-n\right) g\left( nT\right)
\end{equation*}%
\begin{equation}
B\left( t\right) =\sum_{n\in \mathbb{Z}}\left( -1\right) ^{n}\text{sinc}\pi
\left( \frac{t}{T}-n\right) g\left( nT\right) .
\end{equation}%
Both intervals of $\Delta $ are folded on disjoined intervals (a motion of $%
\left( k/T,\left( 2k+1\right) /2T\right) $ towards the left and a motion of
opposite amplitude $\left( -\left( 2k+1\right) /2T,k/T\right) $ towards $%
\left( -1/2T,0\right) ).$ Then, the rhs of $\left( 11\right) $ is reduced to
one term$.$ The hypothesis $k\in \mathbb{N}$ suppresses foldings.\ 

\section{PNS (Periodic Nonuniform Sampling)}

1) The PNS$N$ (Periodic Nonuniform Sampling of order $N$) addresses sampling
sequences%
\begin{equation}
\mathbf{t}=\left\{ nT+\theta _{k},n\in \mathbb{Z}\text{, }k=1,2,..,N\right\}
\end{equation}%
where parameters $\theta _{k}$ are such that the elements of \textbf{t }are
distinct. $\mathbf{t}$ is the concatenation of periodic subsequences%
\begin{equation*}
\mathbf{t}_{k}=\left\{ nT+\theta _{k},n\in \mathbb{Z}\right\}
\end{equation*}%
which are interleaved in \textbf{t. }We already find such cases in \cite{Yen}%
, \cite{Kohl}. From $\left( 4\right) ,$ we deduce the more general formula%
\begin{equation}
e^{2i\pi ft}=\sum_{n=-\infty }^{\infty }e^{2i\pi \alpha T\left( \frac{%
t-\theta }{T}-n\right) }\text{sinc}\left[ \pi \left( \frac{t-\theta }{T}%
-n\right) \right] e^{2i\pi f(nT+\theta )},
\end{equation}%
\begin{equation*}
\alpha -\frac{1}{2T}<f<\alpha +\frac{1}{2T},t\in \mathbb{R}
\end{equation*}%
available for all real $\left( \alpha ,\theta \right) .$ Therefore, we have
iat the same time, for any set $\left( \alpha _{1},\alpha _{2},..,\alpha
_{N},\theta _{1},\theta _{2},..,\theta _{N}\right) $ 
\begin{equation*}
g_{j}\left( t\right) =\int_{\alpha _{j}-1/2T}^{\alpha j+1/2T}G\left(
f\right) e^{2i\pi ft}df
\end{equation*}%
\begin{equation}
g_{j}\left( t\right) =\sum_{n=-\infty }^{\infty }e^{2i\pi \alpha _{j}T\left( 
\frac{t-\theta _{k}}{T}-n\right) }\text{sinc}\left[ \pi \left( \frac{%
t-\theta _{k}}{T}-n\right) \right] g_{j}\left( nT+\theta _{k}\right) .
\end{equation}%
It is the Shannon formula $\left( 5\right) $ in the "frequency band" $\delta
_{j}=\left( \alpha _{j}-\frac{1}{2T},\alpha _{j}+\frac{1}{2T}\right) ,$ for
the periodic sequence \textbf{t}$_{k}$ (delayed by $\theta _{k}$)$.$

2) If we consider the case%
\begin{equation}
g\left( t\right) =\int_{\Delta }G\left( f\right) e^{2i\pi ft}df,\Delta =\cup
_{j=1}^{N}\left( \alpha _{j}-\frac{1}{2T},\alpha _{j}+\frac{1}{2T}\right)
\end{equation}%
we choose the $\alpha _{j}$ so that

\ \ \ \ \ \ \ \ \ \ \ \ \ \ \ \ \ \ \ \ \ 
\begin{equation}
\left\{ \ 
\begin{array}{l}
\alpha _{j+1}-\alpha _{j}\geq 1/T \\ 
\alpha _{j}T\in \beta +\mathbb{Z}\text{.}%
\end{array}%
\right.
\end{equation}%
With the first condition, the $\delta _{k}$ are disjoined. The second one
allows to replace the $g_{j}\left( nT+\theta _{k}\right) $ (unknown
quantities) by their sum $g\left( nT+\theta _{k}\right) $ (what is measured)$%
.$ Therefore, $\left( 15\right) $ yields 
\begin{equation}
\sum_{j=1}^{N}g_{j}\left( t\right) e^{-2i\pi \alpha _{j}\left( t-\theta
_{k}\right) }=\sum_{n=-\infty }^{\infty }e^{-2i\pi n\beta }\text{sinc}\left[
\pi \left( \frac{t-\theta _{k}}{T}-n\right) \right] g\left( nT+\theta
_{k}\right)
\end{equation}%
It is a linear system $N$x$N\,$\ which verifies the matricial equation%
\begin{equation}
\mathbf{MG}_{t}\mathbf{=H}_{t}
\end{equation}%
where the columns matrices $\mathbf{G}_{t},\mathbf{H}_{t}$ and the square
matrix \textbf{M} (independent of $t)$ are defined by%
\begin{equation*}
\mathbf{M=}\left[ e^{2i\pi \alpha _{j}\theta _{k}}\right] ,\mathbf{G}_{t}=%
\left[ g_{j}\left( t\right) e^{-2i\pi \alpha _{j}t}\right]
\end{equation*}%
\begin{equation}
\mathbf{H}_{t}=\left[ \sum_{n=-\infty }^{\infty }e^{-2i\pi n\beta }\text{sinc%
}\left[ \pi \left( \frac{t-\theta _{k}}{T}-n\right) \right] g\left(
nT+\theta _{k}\right) \right]
\end{equation}%
Provided that det\textbf{M}$\neq 0$, which depends on the choice of $\theta
_{k},$ equation $\left( 19\right) $ has a unique solution. We obtain the
value of $\mathbf{G}_{t},$ which leads to the determination of 
\begin{equation}
g\left( t\right) =\sum_{j=1}^{N}g_{j}\left( t\right) .
\end{equation}

3) Let's summarize. We begin by the a periodic sampling and a "baseband
function" ($\Delta =\left( -a,a\right) ).$ Simple extentions provide the
notion of PNS$N$. It is a mixing of a finite number $N$ of periodic
samplings with a common period $T.$ With this model, it is possible to
reconstruct a large choice of functions with "spectral support" $\Delta $
included in the union of $N$ intervals of length $1/T.$ Let's return to the
last example of section 2.2, relaxing hypotheses on $\Delta .$ If 
\begin{equation}
\text{\ \ \ \ }\Delta =\left( \frac{-5}{4T},\frac{-3}{4T}\right) \cup \left( 
\frac{3}{4T},\frac{5}{4T}\right)
\end{equation}%
the periodic sampling $\left\{ nT,n\in \mathbb{Z}\right\} $ is not
sufficient, because both intervals are "folded" on $\left( -1/4T,1/4T\right) 
$ by the (symbolic) operations 
\begin{equation*}
\left( \frac{-5}{4T},\frac{-3}{4T}\right) +\frac{1}{T}=\left( \frac{3}{4T},%
\frac{5}{4T}\right) -\frac{1}{T}
\end{equation*}%
Nevertheless, for $a\notin T\mathbb{Z}$/$2,$ we prove that%
\begin{equation*}
g\left( t\right) =\frac{1}{\sin 2\pi a/T}\left[ -A_{0}\left( t\right) \sin
2\pi \left( t-a\right) /T+A_{a}\left( t\right) \sin 2\pi t/T\right]
\end{equation*}%
\begin{equation}
A_{x}\left( t\right) =\sum_{n=-\infty }^{\infty }\text{sinc}\left[ \pi
\left( \frac{t-x}{2T}-n\right) \right] g\left( 2nT+x\right) .
\end{equation}%
Therefore, with the PNS$2,$ \textbf{t}=$\left\{ 2nT,2nT+a,n\in \mathbb{Z}%
\right\} $, we cancel the drawback of a folding which appears in a periodic
sampling \cite{Kohl}, \cite{Jerr}, \cite{Laca8}. The property is generalized
in: with a well-chosen PNS$N,$ it is possible to overcome $N-1$ foldings.

\section{Nyquist bound and Landau bound}

To linearly reconstruct 
\begin{equation}
g\left( t\right) =\int_{\Delta }G\left( f\right) e^{2i\pi ft}df
\end{equation}%
from the sampling sequence \textbf{t}=$\left\{ t_{n},n\in \mathbb{Z}\right\}
,$ is equivalent to find a formula as%
\begin{equation}
e^{2i\pi ft}=\sum_{n=-\infty }^{\infty }a_{n}\left( t\right) e^{2i\pi
ft_{n}},f\in \Delta ,t\in \mathbb{R}
\end{equation}%
because, taking some caution, we obtain from $\left( 24\right) $%
\begin{equation}
g\left( t\right) =\sum_{n=-\infty }^{\infty }a_{n}\left( t\right) g\left(
t_{n}\right) .
\end{equation}%
The "Nyquist condition" is summarized by 
\begin{equation}
\Delta =\left( -\frac{1}{2\mu _{0}},\frac{1}{2\mu _{0}}\right)
\Longrightarrow \lim_{\left\vert n\right\vert \rightarrow \infty }\frac{t_{n}%
}{n}=\mu \leq \mu _{0}.
\end{equation}%
It is a necessary condition for reconstruction: if the sampling sequence
verifies $\lim_{\left\vert n\right\vert \rightarrow \infty }\frac{t_{n}}{n}%
=\mu $ for some finite $\mu ,$ a necessary condition for errorless
reconstruction of any $g\left( t\right) $ fulfilling $\left( 24\right) $ is $%
\mu \leq \mu _{0}.$ It is not sufficient (see section 5.1), and the main
drawback of this condition comes from the shape of $\Delta .$ The Nyquist
condition answers problems of analysis about the completion of complex
exponentials \cite{Levinson}, and sufficient conditions can be found in
particular cases (Kadec's th. for example in case of "jitter"). The
condition $\left( 27\right) $ becomes very strong when $\mu _{0}$ is small,
and physical models (particularly in communications) prescribe other shapes
for $\Delta .$

The "Landau condition" \cite{Land} is summarized by%
\begin{equation}
\Delta =\cup _{k}\left( a_{k},b_{k}\right) \Longrightarrow \lim_{\left\vert
n\right\vert \rightarrow \infty }\frac{t_{n}}{n}=\mu \leq \frac{1}{%
\left\vert \Delta \right\vert }.
\end{equation}%
For an errorless reconstruction when $\Delta $ is an union of intervals, the
"sampling rate" $\mu $ has to be smaller (or equal) to the length of $\Delta
.$ Landau condition is weaker than Nyquist condition, and better suited. For
instance, we know that the frequency spectrum used in communications is
divided in bands corresponding to telephony, television... This gives a
particular interest to the case%
\begin{equation*}
\Delta =\left( -b,-a\right) \cup \left( a,b\right) .
\end{equation*}%
In section 2.1 we consider the simplest situation (a Nyquist one) 
\begin{equation*}
t_{n}=nT,\mu =\mu _{0}=T.
\end{equation*}%
In section 2.2, examples refer to Landau condition, and errorless
reconstruction.

\section{Irregular sampling}

\subsection{Lagrange interpolation}

The Lagrange interpolation sums up points clouds by polynomials \cite{Lagr}, 
\cite{Wari}. The extension to entire functions is a part of the complex
analysis \cite{Levi}. Being given the increasing sequence \textbf{t}=$%
\left\{ t_{n},n\in \mathbb{Z}\right\} ,$ and $g\left( t\right) $ in the form 
$\left( 2\right) $ for instance, the Lagrange interpolation brings up the
validity problem of the equality (in some sense)

\begin{equation*}
g\left( t\right) =\sum_{n\in \mathbb{Z}}\frac{H\left( t\right) }{\left(
t-t_{n}\right) H^{\prime }\left( t_{n}\right) }g\left( t_{n}\right)
\end{equation*}%
\begin{equation}
\text{with :\ \ \ \ \ \ \ \ }H\left( t\right) =\dprod\limits_{n\in \mathbb{Z}%
}\left( 1-\frac{t}{t_{n}}\right) \text{ or }H\left( t\right)
=t\dprod\limits_{n\in \mathbb{Z}^{\ast }}\left( 1-\frac{t}{t_{n}}\right)
\end{equation}%
following that $0\notin \mathbf{t}$ or $t_{0}=0\in \mathbf{t.}$ In baseband, 
$\Delta =\left( -1/2T,1/2T\right) ,$ the Nyquist condition has to be
verified \cite{Levi}, but it is not sufficient. For instance $\left(
29\right) $ is true for \textbf{t=}$\mathbb{Z}$, $T=1$. It is the standard
formula knowing that

\begin{equation}
\sin \pi x=\pi x\dprod\limits_{n\neq 0}\left( 1-\frac{x}{n}\right) .
\end{equation}%
Formula $\left( 29\right) $ is not verified for \textbf{t=}$\mathbb{Z}$-$%
\left\{ 0\right\} ,$ even if $G\left( f\right) $ cancels at the neighbourood
of $\pm 1/2$ (to increase the Nyquist bound$).$ In this case, it is easy to
find alternative formulas. The problem is linked to the behavior of $H\left(
t\right) $ at the infinite points. The difference comes from

\begin{equation*}
\overline{\lim }_{\left\vert x\right\vert \rightarrow \infty }\left\vert
\sin x\right\vert =1\text{ vs lim}_{\left\vert x\right\vert \rightarrow
\infty }\frac{\sin x}{x}=0
\end{equation*}%
where $t_{0}=0$ (\textbf{t}=$\mathbb{Z}$) or not (\textbf{t=}$\mathbb{Z}$-$%
\left\{ 0\right\} )$. In real situations, devices compute a reduction $%
\widetilde{g}\left( t\right) $ of $\left( 29\right) $ 
\begin{equation}
\widetilde{g}\left( t\right) =\sum_{n=\left[ t\right] -N}^{\left[ t\right]
+N}\frac{\widetilde{H}\left( t\right) }{\left( t-t_{n}\right) \widetilde{H}%
^{\prime }\left( t_{n}\right) }g\left( t_{n}\right)
\end{equation}%
because of limited capacities of devices and the future of $t_{n}$ is not
very well known ($\left[ t\right] $ is close to $t$ so that it is the
neighbourhood of $t$ which is taken into account). This type of
approximation enters the class of operations named "\textbf{apodization}". $%
\widetilde{H}\left( t\right) $ can take many shapes. The simplest is to
retain the $t_{n}$ in some interval around $t$ \cite{Laca1}, \cite{Laca6}$.$

Another widely used technique is the "\textbf{resampling"}. Let's assume
that formula $\left( 5\right) $ is available and that we know $g\left(
t_{k}\right) $ in a neighbourhood of $t.$ For $N$ large enough, we will
often have 
\begin{equation*}
g\left( t_{k}\right) \approx \sum_{n=\left[ t\right] -N}^{\left[ t\right] +N}%
\text{sinc}\left[ \pi \left( \frac{t_{k}}{T}-n\right) \right] g\left(
nT\right) ,\left\vert k\right\vert \leq N.
\end{equation*}%
This (approximative) linear system $\left( N+1\right) $x$\left( N+1\right) $
provides the $g\left( nT\right) $ which allows to determine estimations of $%
g\left( t\right) $ whatever $t$.

\subsection{A solution from PNS}

1) We consider the (increasing) sampling sequence \textbf{t}=$\left\{
t_{n},n\in \mathbb{Z}\right\} ,$ and a function $g\left( t\right) $
verifying $\left( 16\right) ,$ $\left( 17\right) $. We have $\left\vert
\Delta \right\vert =N/T,$ and we take%
\begin{equation}
\mathbf{t}^{N_{0}}=\left\{ nT+\theta _{k},n\in \mathbb{Z}\text{, }%
k=1,2,..,N\right\} ,\theta _{k}=t_{N_{0}+k}.
\end{equation}%
$\left( 18\right) ,\left( 19\right) $ are true whatever integer $N_{0}$.
Let's assume that $t_{n+1}-t_{n}<T/N,$ so that the sequence $\mathbf{t}%
^{N_{0}}$ shows a natural order due to the inequality%
\begin{equation*}
t_{N_{0}+N}-T<t_{N_{0}+1}<...<t_{N_{0}+N}<t_{N_{0}+1}+T.
\end{equation*}%
For a given $t,$ we choose $N_{0}=N_{0}\left( t\right) $ such that (for even 
$N)$ 
\begin{equation*}
t_{N_{0}+1}<...<t_{N_{0}+N/2}\leq t\leq t_{N_{0}+1+N/2}<...<t_{N_{0}+N}.
\end{equation*}%
In such a situation, when $N$ is large enough, we can replace $\left(
19\right) ,$ $\left( 20\right) $ by the approximations%
\begin{equation*}
\mathbf{M}\widetilde{\mathbf{G}}_{t}\mathbf{=}\widetilde{\mathbf{H}}_{t}
\end{equation*}%
\begin{equation*}
\mathbf{M=}\left[ e^{2i\pi \alpha _{j}\theta _{k}}\right] ,\widetilde{%
\mathbf{G}}_{t}=\left[ \widetilde{g}_{j}\left( t\right) e^{-2i\pi \alpha
_{j}t}\right] ,
\end{equation*}%
\begin{equation}
\widetilde{\mathbf{H}}_{t}=\left[ \text{sinc}\left[ \pi \left( \frac{%
t-\theta _{k}}{T}\right) \right] g\left( \theta _{k}\right) \right] .
\end{equation}%
This means that we suppress in \textbf{H}$_{t}$ the terms $n\neq 0,$ which
have abscissas the farthest away from $t.$ These terms are not measured. The
estimation of $g\left( t\right) $ is obtained from $N$ measures $g\left(
t_{k}\right) ,$ at points $t_{j}$ which are the closest to $t.$ It is
equivalent to an apodization of an errorless formula, not taking into
account unknown quantities (the $g\left( nT+\theta _{k}\right) $ which are
not measured for $n\neq 0).$

2) For each $\left( N,N_{0}\right) ,$ the sequence $\mathbf{t}^{N_{0}}$
defined in $\left( 32\right) $ verifies the Landau condition, even if lim$%
_{n\longrightarrow \pm \infty }t_{n}/n$ does not exist. The quality of
estimations $\widetilde{\mathbf{G}}_{t}$ of $\mathbf{G}_{t}$ depends on the
"distance" between $\widetilde{\mathbf{H}}_{t}$ and $\mathbf{H}_{t}$.
Obviously, the technique is well suited when $t$ is chosen in the interval $%
\left( t_{1},t_{N}\right) ,$ and far from the $nT+t_{k},n\neq 0,$ where we
can expect that the omitted terms are negligible.

When $G\left( f\right) $ is regular enough and $N$ large enough, so that the
mean theorem of integral can be involved, we have, from (15$)$%
\begin{equation}
g_{k}\left( t\right) \approx \frac{\left\vert \Delta \right\vert }{N}e^{i\pi
\alpha _{k}t}G\left( \alpha _{k}\right) .
\end{equation}%
Approximately, the "spectral content " of $g\left( t\right) $ at $\alpha
_{k} $ is close to $\frac{\left\vert \Delta \right\vert }{N}G\left( \alpha
_{k}\right) .$ Equivalently, the filtering of $g\left( t\right) $ in the
band $\delta _{k}$ is quasi-monochromatic, its amplitude and its phase being
defined by $\frac{\left\vert \Delta \right\vert }{N}G\left( \alpha
_{k}\right) .$

\section{Stationary processes}

\textbf{Z=}$\left\{ Z\left( t\right) ,t\in \mathbb{R}\right\} $ is a (wide
sense) stationary random process when 
\begin{equation}
K\left( \tau \right) =\text{E}\left[ Z\left( t\right) Z^{\ast }\left( t-\tau
\right) \right] =\int_{\Delta }e^{2i\pi f\tau }s\left( f\right) df
\end{equation}%
where E$\left[ ..\right] $ is the mathematical expectation (ensemble mean), $%
a^{\ast }$ is the complex conjugate of $a$ \cite{Papo}, \cite{Cram}. The
integral has the common sense (a Riemann integral) when $s\left( f\right) $
is regular enough. Then, $s\left( f\right) $ is real and positive ($\geq 0)$
and is symmetric for real \textbf{Z.} Elsewhere, $s\left( f\right) $ is
explained by $\left( 6\right) ,$ $\left( 7\right) .$ It is a sum of two
terms, the first one is an ordinary function (but positive) the other one is
a sum of complex exponentials with positive coefficients. $s\left( f\right) $
defines the "power spectrum" of \textbf{Z.} $\Delta $ is a support of the
spectrum ($s\left( f\right) =0$ outside $\Delta ).$

The sampling problem is formulated in the same way as for ordinary functions 
\cite{Laca4}. From the $\left( t_{n},Z\left( t_{n}\right) \right) ~$we have
to find a linear reconstruction of $Z\left( t\right) ,$ i.e a formula as

\begin{equation}
Z\left( t\right) =\sum_{n\in \mathbb{Z}}a_{n}\left( t\right) Z\left(
t_{n}\right)
\end{equation}%
where $a_{n}\left( t\right) $ depends only of $\Delta $ (and not of values
of $s\left( f\right) $ on $\Delta ).$ In the mean-square sense, equality $%
\left( 35\right) $ verifies%
\begin{equation*}
E\left[ \left\vert Z\left( t\right) -\sum_{n\in \mathbb{Z}}a_{n}\left(
t\right) Z\left( t_{n}\right) \right\vert ^{2}\right] =0\Leftrightarrow
\end{equation*}%
\begin{equation*}
\int_{\Delta }\left\vert e^{2i\pi f\tau }-\sum_{n\in \mathbb{Z}}a_{n}\left(
t\right) e^{2i\pi ft_{n}}\right\vert ^{2}s\left( f\right) df=0
\end{equation*}%
This means that we have to develop $e^{2i\pi ft}$ on $\Delta $ as function
of $e^{2i\pi ft_{n}},n\in \mathbb{Z}$. With mathematical nuances about
differences between kinds of convergences, integrals..., the sampling
problem remains the same that for functions like $\left( 2\right) $ which
have Fourier transforms on some set $\Delta .$ We have to look for a
development of $e^{2i\pi ft}$ on $\Delta $ as

\begin{equation}
e^{2i\pi ft}=\sum_{n\in \mathbb{Z}}a_{n}\left( t\right) e^{2i\pi
ft_{n}},f\in \Delta ,t\in \mathbb{R}
\end{equation}%
and to replace $e^{2i\pi ft}$ by $Z\left( t\right) ,$ for $t,t_{j},j\in 
\mathbb{Z}$, in $\left( 35\right) .$ We obtain, from section 3 $\ $ 
\begin{equation*}
\mathbf{M}\widetilde{\mathbf{G}}_{t}=\widetilde{\mathbf{H}}_{t}
\end{equation*}%
\begin{equation*}
\mathbf{M=}\left[ e^{2i\pi \alpha _{j}t_{k}}\right] ,\widetilde{\mathbf{G}}%
_{t}=\left[ \widetilde{Z}_{j}\left( t\right) e^{-2i\pi \alpha _{j}t}\right]
\end{equation*}%
\begin{equation}
\widetilde{\mathbf{H}}_{t}=\left[ \text{sinc}\left[ \pi \left( \frac{t-t_{j}%
}{T}\right) \right] Z\left( t_{j}\right) \right]
\end{equation}%
with the approximation%
\begin{equation}
\widetilde{Z}\left( t\right) =\sum_{k=1}^{N}\widetilde{Z}_{k}\left( t\right)
\end{equation}%
$Z_{k}\left( t\right) $ reults from a linear filtering of $\mathbf{Z}$ on $%
\delta _{k}.$ The \textbf{Z}$_{k}$ are uncorrelated (the $\delta _{k}$ are
disjoined) and%
\begin{equation}
K_{k}\left( \tau \right) =\text{E}\left[ Z_{k}\left( t\right) Z_{k}^{\ast
}\left( t-\tau \right) \right] =\int_{\delta _{k}}e^{2i\pi f\tau }s\left(
f\right) df
\end{equation}%
In $\left( 34\right) ,$ the power spectrum $s\left( f\right) $ (for \textbf{Z%
}) has the sense attributed to $G\left( f\right) $ \ (for $g\left( t\right)
) $ in section 2.1. $s\left( f\right) $ is the sum of a continuous part and
a discrete part. This means that we observe accumulations of power in some
places, which are preserved when $N$ increases.

\section{Power spectra measurements}

From $\left( 34\right) $, the power spectrum $s\left( f\right) $ of \textbf{%
Z,} is the inverse Fourier transform of the correlation function $K\left(
t\right) ,$ classically estimated from%
\begin{equation}
\widetilde{K}\left( nT\right) =\frac{1}{N-n}\sum_{k=1}^{N-n}Z\left(
kT\right) Z\left( \left( k+n\right) T\right) ,n=0,..,N-1
\end{equation}%
in the case of a periodic sampling$.$ The estimation worsens when $n$
increases. Generally, people consider that the accuracy in the research of
high frequencies (v.s low) is very dependent on the accuracy in the
estimation for low values (v.s high) of $n.$ If true, we have to dispose of
samples in a large interval of time if we want low frequencies, and samples
very close for high frequencies. Moreover, in the case of irregular
sampling, $\left( 40\right) $ will be achevied when the $Z\left( kT\right) $
will be themselves estimated from the $Z\left( t_{k}\right) $.

When integer $N$ is large enough, $s\left( f\right) $ can be estimated
directly from PNS. To simplify notations, we take the particular case%
\begin{equation*}
\Delta =\left( -b,b\right) ,\delta _{k}=\left( -b+\frac{2b}{N}\left(
k-1\right) ,-b+\frac{2b}{N}k\right) ,k=1,2,..,N.
\end{equation*}%
We have $\left\vert \delta _{k}\right\vert =2b/N$. Let's admit that $\left(
37\right) $ provides good estimations of elements of $\mathbf{G}_{t}$ 
\begin{equation*}
Z_{k}\left( t\right) e^{-2i\pi \alpha _{k}t},\alpha _{k}=-b+\frac{b}{N}%
\left( 2k-1\right)
\end{equation*}%
Each term of $\mathbf{G}_{t}$ belongs to the frequential set $\left(
-b/N,b/N\right) .$ This means that $Z_{j}\left( t\right) e^{-2i\pi \alpha
_{j}t}$ is a demodulation of $Z_{j}\left( t\right) $ in baseband. When $b/N$
is small enough, $Z_{k}\left( t\right) $ appears as the monochromatic wave $%
a\left( t\right) e^{2i\pi \alpha _{k}t}$ with an amplitude $\left\vert
a\left( t\right) \right\vert $ which varies slowly. The total power $P_{k}$
of \textbf{Z}$_{k}$ is (it is a definition)%
\begin{equation*}
P_{k}=\text{E}\left[ \left\vert Z_{k}\left( t\right) \right\vert ^{2}\right]
=\text{E}\left[ \left\vert G_{k}\left( t\right) \right\vert ^{2}\right]
\end{equation*}%
which is estimated measuring the mean of $\left\vert G_{k}\left( t\right)
\right\vert ^{2}$ in $\left( 0,t_{N}\right) .$

\section{Example: the Vostok ice core}

\subsection{The evolution of CO$_{2}$}

Vostok ice cores are columns of ice of near four kilometres length coming
from the top to the bottom of a lake of southern antarctica (near the
magnetic south pole). The top (v.s the bottom) corresponds to recent (v.s
around 400000 years old) snow. The study of samples collected along the ice
core gives many indications, the date, temperature, composition and
variations of elements as CO$_{2},$ CH$_{4},$ D$_{2}$ (deuterium), O$^{18}$
(isotope of oxygen)... Figure 1 shows the evolution of CO$_{2}$
concentration on a period of 400000 years (the past at the right). Data are
in * and interpolation in - - -. Figure 2 shows the corresponding power
spectrum on $\left( 0,7.10^{-5}\right) $ in yr$^{-1},$ with $N=282$ and $%
\left\vert \delta _{k}\right\vert =7.10^{-7}$yr$^{-1}.$ The original
estimation is in inset \cite{Peti}. Improvements on the accuracy of
estimations are clear, by the sharpeness of spectral rays and by the
appearance of low frequency components. Figure 3 shows components 3, 9,
10,... ($Z_{3}\left( t\right) ,Z_{9}\left( t\right) ..).$ In figure 4 we see
an addition of the first three rays in low frequencies. Figure 5 represents
the power spectrum of CH$_{4}.$ It is very similar to the former one (figure
2), with the very strong ray at 1$0^{-5}$yr$^{-1}$ (related to the 100000 yr
problem \cite{Rial}).

The Vostok ice core is a very telling example of irregular sampling. Data
are not in arithmetic series, due to varying conditions of precipitation and
mechanical loads \cite{Laca9}, \cite{Bona}. Estimations of power spectra
have to be compared to results in \cite{Babu}, where are detailed the best
methods by the best specialists.

\subsection{Remark}

In the Vostok ice core, we have taken $\Delta =$ $\left(
-10^{-4},10^{-4}\right) $ and the $\delta _{k}$ are adjacent. Nyquist and
Landau conditions are equivalent. It is no longer true when $\Delta $ is
piecewise. Figure 6 addresses a Gaussian process in baseband modulated in $%
\left( -5/2,-2\right) \cup \left( 2,5/2\right) $ i.e multiplied by cos $%
\left( 9\pi t/2\right) .$ Sampling times are in the form

\begin{equation*}
t_{n}=n\left( 1-\varepsilon \right) +A_{n},0<\varepsilon <1/2
\end{equation*}%
where the $A_{n}$ are uniformly distributed in $\left( -1/2,-1/4\right) \cup
\left( 1/4,1/2\right) .$ The estimation at time $t$ is done using 10 samples
before $t$ and 10 after. On figure 6, we find samples (*) and a perfect
reconstruction (at the thickness of the line). We have%
\begin{equation*}
\lim_{n\rightarrow \infty }\frac{t_{n}}{n}=1-\varepsilon <1
\end{equation*}%
which verifies the Landau condition, but not the Nyquist one $\left(
\lim_{n\rightarrow \infty }t_{n}/n=5\right) .$ This shows that the Nyquist
condition is often pessimistic.

\section{Conclusion}

This paper addresses irregular sampling of ordinary functions $g\left(
t\right) $ and random processes \textbf{Z }with bounded support spectrum. We
explain why PNS (Periodic Nonuniform Samplings) are the right framework when
we have at our disposal a large enough number of samples (but a finite
number). To construct errorless formulas, we begin by the periodic sampling,
suited to an arbitrary interval. This leads to fit PNS$N$ to piecewise power
spectra, verifying the Landau condition ($N$ is the number of samples). The
reconstruction of $g\left( t\right) $ or \textbf{Z }is done through a $N$x$N$
matricial system which solutions are the $N$ components corresponding to
pieces of power spectra. The mathematical context does not exceed the
elementary framework of Fourier analysis (one-dimensional series and
integrals).

Considerations about Vostok ice cores illustrate theoretical results. They
provide data which give strong informations about climate in the last 4.10$%
^{5}$ years. Physical-chemical studies have allowed to measure the
concentration in C0$_{2},$ CH$_{4},$ isotopes of oxygene... of atmosphere
trapped in ice at irregular times. Sampling methods explained here allow to
reconstitute these quantities and their evolution on a long period of the
past. The proposed method, thanks to band decomposition in small intervals,
shows improvements in the number and the thickness of spectral rays.
Particularly, low frequency components appear, which are hidden in usual
computations.

\section{Appendix: about Fourier series (FSD)}

1) Let's $h\left( f\right) $ such that%
\begin{equation*}
h\left( f\right) =h\left( f+f_{0}\right)
\end{equation*}%
for "$h\left( f\right) $ is periodic with period $f_{0}".$ Fourier
cefficients $c_{n}$ are defined as 
\begin{equation*}
c_{n}=\frac{1}{T}\int_{-f_{0}/2}^{f_{0}/2}h\left( f\right) e^{-2i\pi
nf/f_{0}}df.
\end{equation*}%
We have the following property (provided some conditions about the $h\left(
f\right) $ regularity and the convergence definition$).$%
\begin{equation}
\text{If }\widetilde{h}\left( f\right) =\sum_{n=-\infty }^{\infty
}c_{n}e^{2i\pi nf/f_{0}},\text{ then }h\left( f\right) =\widetilde{h}\left(
f\right) .
\end{equation}%
$\widetilde{h}\left( f\right) $ is the Fourier Series Development (FSD) of $%
h\left( f\right) .$

2) This definition of FSD is dangerous. Firstly, a function $h\left(
f\right) $ being given, we may look for a development of $h\left( f\right) $
verifying $\left( 42\right) $ on some set $\Delta $ for non-periodic $%
h\left( f\right) $.

For instance, if $h\left( f\right) =f$ ,$f\in \mathbb{R}$, $\Delta =\left(
a,a+b\right) ,$ we have%
\begin{equation*}
c_{n}=\frac{1}{b}\int_{a}^{a+b}fe^{-2i\pi nf/b}df=\frac{-b}{2i\pi n}%
e^{-2i\pi na/b},n\neq 0
\end{equation*}%
\begin{equation*}
\widetilde{h}\left( f\right) =a+\frac{b}{2}-\frac{b}{\pi }\sum_{n=1}^{\infty
}\frac{1}{n}\sin \left[ \frac{2\pi n}{b}\left( f-a\right) \right] .
\end{equation*}%
For $n\in \mathbb{Z}$ and except at the bounds of intervals : 
\begin{equation*}
\widetilde{h}\left( f\right) =\left\{ 
\begin{array}{c}
h\left( f\right) ,f\in \left( a,a+b\right) \\ 
h\left( f\right) -nb,f\in \left( a+nb,a+\left( n+1\right) b\right)%
\end{array}%
\right.
\end{equation*}%
Figure 7 explains the result. We have found a trigonometric development of $%
h\left( f\right) $ on $\Delta $ (it is $\widetilde{h}\left( f\right) ).$ $%
\widetilde{h}\left( f\right) $ is periodic, as a sinus, and then is
different from $h\left( f\right) $ outside $\Delta .$ Changing the value of $%
a$ and/or $b$ yields a FSD different on another set $\Delta .$

3) Then, a function defined on the whole $\mathbb{R}$, provides a different
FSD following the set $\Delta $ where the development takes place. It is
true when $h\left( f\right) $ is periodic with period $f_{0}.$ Obviously, $%
h\left( f\right) $ can be developped on intervals of length different of $%
f_{0}.$ In the beginning of this paper, we prove the standard sampling
formula (Shannon, Nyquist,...) using FSD when 
\begin{equation*}
\Delta =\left( -1/2T,1/2T\right) \ \text{and }h\left( f\right) =e^{2i\pi ft}.
\end{equation*}%
$h\left( f\right) $ is a periodic function of $f\in \mathbb{R}$ with period $%
f_{0}=1/t.$ Formula $\left( ..\right) $ is the DSF of $h\left( f\right) $ on 
$\Delta .$ It is natural that coefficients depend on parameter $t,$ but $t$
it is not the used variable. The obtained DSF provides $h\left( f\right) $
on $\Delta $ but not elsewhere (except the particular case $t=T).$ Figure 8
illustrates this very fundamental situation.

\end{document}